\documentclass[prc,aps,10pt,twocolumn,showkeywords,showpreprintnumbers,floatfix,nofootinbib]{revtex4-1}
\usepackage{amsmath,amssymb,graphicx,bm}
\usepackage{bbm}
\usepackage{hyperref}
\usepackage{color,ulem}
\usepackage{slashed}
\allowdisplaybreaks[4]

\newcommand{\vect}[1]{\bm{\mathrm{#1}}}

\newcommand{\qv}{\vect{q}}

\newcommand{\fm}{\;\text{fm}}
\newcommand{\MeV}{\;\text{MeV}}

\newcommand{\sv}{\vect{s}}
\newcommand{\nablav}{\vect{\nabla}}
\newcommand{\laplacian}{\Delta}

\definecolor{RED}{rgb}{1,0,0}

\DeclareMathOperator{\Imag}{Im}
\renewcommand{\Im}{\Imag}

\renewcommand{\emph}[1]{\textit{#1}}
\newcommand{\iso}{I}

\begin{document}

\title{New Skyrme parametrizations to describe finite nuclei and
  neutron star matter with realistic effective masses. II. Adjusting
  the spin-dependent terms}

\author{Mingya Duan}\email{mingya.duan@ijclab.in2p3.fr}
\affiliation{Universit\'e
  Paris-Saclay, CNRS/IN2P3, IJCLab, 91405 Orsay, France}

\author{Michael Urban} \email{michael.urban@ijclab.in2p3.fr}
\affiliation{Universit\'e
  Paris-Saclay, CNRS/IN2P3, IJCLab, 91405 Orsay, France}

\begin{abstract}
Many common Skyrme functionals present ferromagnetic instabilities or
unrealistic density dependence of the spin-dependent Landau parameters. To
solve these problems, we consider the Skyrme interaction as a
density-functional rather than a density-dependent two-body
force. This allows us to adjust the spin-dependent terms of the new
extended Skyrme functionals of our previous paper [M. Duan and
  M. Urban, Phys. Rev. C 110, 065806 (2024)] independently without
altering the properties of spin saturated matter. The parameters of
the spin-dependent terms are determined by fitting the Landau
parameters $G_0$ and $G_0^{\prime}$ in neutron matter and symmetric
nuclear matter and the effective-mass splitting of up and down
particles in spin polarized matter to the results of microscopic
calculations. Using the new parametrizations, called Sky3s and Sky4s,
the spin-related properties of nuclear matter are in good agreement
with the microscopic results. As an application, we compute response
functions and neutrino scattering rates of neutron-star matter with
the new functionals having realistic effective masses and Landau
parameters.
\end{abstract}

\maketitle

\section{Introduction}\label{sec:introduction}
Skyrme effective interactions are easy to use due to their zero-range
character. The energy density functional derived from Skyrme
interactions has been widely used to study nuclear structure and
neutron star properties
\citep{Vautherin1972,Beiner1975,Kohler1976,Chabanat1997,Goriely2010}. To
compute neutron star properties, interaction SLy4, constructed in 1998
\citep{Chabanat1998}, is very popular. Since 2010, a series of BSk
interactions have also been developed for astrophysical applications
\citep{Goriely2010,Goriely2013,Goriely2016}.

An important application of the nuclear energy density functional is
to describe the transport properties of neutrinos in (proto-)neutron
star and supernova matter. Neutrinos play a crucial role in explaining
supernova explosions, neutron star mergers, proto-neutron star
evolution, and neutron star cooling
\citep{Hirata1987,Mezzacappa2005,Janka2012,Burrows2013,Burrows1986,Mayle1987,Keil1994,Pascal2022,Yakovlev2004}.
These astrophysical processes occur along with neutrino absorption and
scattering. As many studies have done, such as the early literature
\cite{Iwamoto1982} and the recent references
\cite{Pastore2015,Oertel2020,Duan2023,Lin2023}, the neutrino rates (absorption
and scattering rates) or the neutrino mean free path can be related to
the response functions, including responses computed using the Skyrme
interactions.

\begin{figure}
\begin{center}
\includegraphics[scale=0.5]{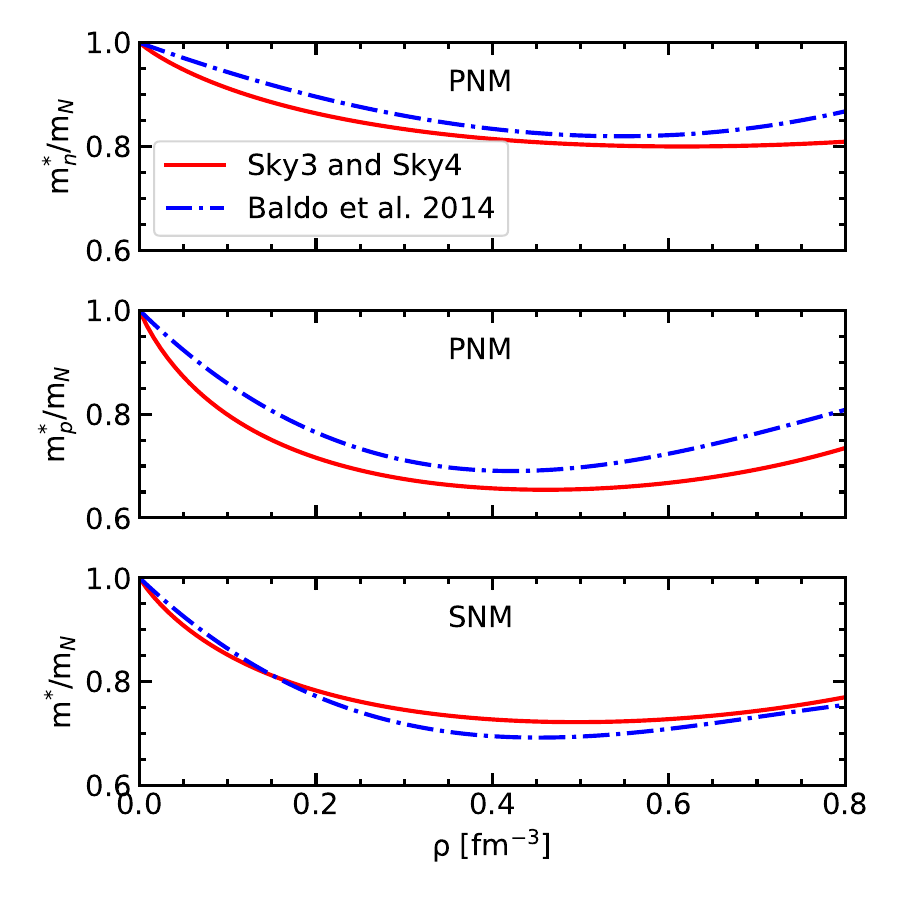} 
\caption{(Top) Neutron effective mass in PNM, (middle) proton
  effective mass in PNM (= isovector effective mass), and (bottom)
  nucleon effective mass in SNM (= isoscalar effective mass) from the
  BHF calculations of Ref. \cite{Baldo2014} (dash-dot lines) and the
  corresponding ones of Sky3 and Sky4 \cite{Duan2024} (solid lines).}
\label{fig:effective-masses-PNM-SNM}
\end{center}
\end{figure}

A couple of articles have studied neutrino mean free path using Skyrme
energy density functional for pure neutron matter
\citep{Pastore2012,Pastore2014,Pastore2015} and asymmetric nuclear
matter \citep{Davesne2014,Davesne2019}. But we recently reported that
many Skyrme effective interactions predict an unrealistic density
dependence of the effective masses, resulting, e.g., in the unphysical
feature that the neutron Fermi velocity exceeds the speed of light at
relatively low densities \citep{Duan2023}. Therefore new
parametrizations were needed to solve this problem. We constructed two
new Skyrme interactions, Sky3 and Sky4, solving the problem by
adjusting the parameters to Brueckner-Hartree-Fock (BHF) results for
the effective masses \citep{Duan2024}. Figure
\ref{fig:effective-masses-PNM-SNM} displays the effective masses in
pure neutron matter (PNM) and in symmetric nuclear matter (SNM) from
the BHF calculations of Ref. \cite{Baldo2014} and the corresponding
ones of Sky3 (which are identical to those of Sky4). We can see that
the neutron effective mass in PNM and the nucleon effective mass in
SNM computed with Sky3 and Sky4, which will be used in the
calculations and fits in the present paper, are in good agreement with
the BHF ones, although some discrepancies cannot be avoided, mainly
because the relation
$1/m_p^{*}(\text{PNM})+1/m_n^{*}(\text{PNM})=2/m^{*}(\text{SNM})$
imposed by Skyrme functionals is not satisfied by the BHF effective
masses \cite{Duan2024}.

However, there is still a problem in these two interactions, namely,
they predict a ferromagnetic instability at densities that may be
realized in neutron stars \citep{Duan2024}. The reason is that the
spin dependent terms of the functionals are expressed in terms of the
same parameters as the spin-independent ones, and the fitting focuses
on observables that depend only on the spin-independent terms with the
exception of the spin-orbit parameter $W_0$. Apart from the unphysical
instability, the spin dependent terms affect also the neutrino
scattering rates which involve contributions of spin-0
(spin-independent) and spin-1 (spin-dependent) channels. The response
functions of neutron-star and supernova matter, thus the neutrino
rates, can be computed only if the Landau parameters in the spin-1
channel have the correct values. Therefore, the spin-dependent terms
of the new Skyrme functionals should be constrained reasonably.

The spin Landau parameters can be related to spin- and spin-isospin
excitations, such as magnetic dipole (M1) and Gamow-Teller (GT)
resonances.  Already more than 40 years ago, the spin dependent Landau
parameters at saturation density have been obtained from fits to
GT-resonance systematics
\cite{Bertsch1981a,Gaarde1981,Suzuki1982,Bertsch1981b}. Research on
this topic has been pursued since then
\cite{Bender2002,Wakasa2005,Wakasa2012,Yasuda2018}.  Thus, a
possibility to constrain the spin properties of Skyrme functionals
(especially in the isovector channel) was to include data for
Gamow-Teller resonances \cite{Osterfeld1992,Ichimura2006} into the
fitting protocol. Examples for interactions constrained in this way
are SGII \cite{VanGiai1981} and SAMi \cite{Roca-Maza2012}. However, in
these interactions the spin dependent and the spin independent terms
are linked to each other, since they are all determined by the
standard Skyrme parameters. A different approach was taken in
Ref. \cite{Bender2002}. There, the Skyrme interaction was interpreted
as a density functional instead of a density-dependent two-body force,
which allowed the authors to adjust the spin dependent terms
independently to describe the Gamow-Teller strength distribution,
without changing the other terms (which in that work were taken from
the SkO$^{\prime}$ parametrization \cite{Reinhard1999}). In the
present work, we will also adjust the spin dependent terms
independently, but instead of fitting them to Gamow-Teller resonances,
we will adjust them to the results of microscopic calculations
\cite{Zuo2003communications,Zuo2003PNM,Zuo2003SNM,Bigdeli2009,Vidana2016,Bigdeli2024,Vidana2024},
similar to earlier attempts to improve the spin properties of the SLy5
and BSk17 interactions \cite{Margueron2009,Margueron2009b}.

In Sec. \ref{sec:new Skyrme EDF}, we first point out the problems of
the spin-related properties of nuclear matter with the unconstrained
Skyrme functionals Sky3 and Sky4, but also with other Skyrme
functionals. Then, we present the strategies to adjust the
spin-dependent terms of the functionals by using results of BHF
calculations in Sec. \ref{sec:constraints to the spin dependent
  terms}. The Landau parameters in pure neutron matter and symmetric
nuclear matter, the equation of state of spin-polarized neutron
matter, and the instabilities in neutron star matter computed with the
resulting new functionals Sky3s and Sky4s are also shown in
Sec. \ref{sec:constraints to the spin dependent terms}. Finally, we
present some results of the full RPA (random phase approximation)
response functions of neutron star matter and the corresponding
neutrino scattering rates computed with Sky3s in
Sec. \ref{sec:responses-neutrino-rates}, and summarize in
Sec. \ref{sec:conclusion}.

\section{Spin-related properties of nuclear matter with Skyrme functionals} 
\label{sec:new Skyrme EDF}
The standard form of the Skyrme energy density functional consists of
the kinetic-energy term, the zero-range term, the density-dependent
term, the effective-mass term, the finite-range term, the spin-orbit
term, the tensor coupling term, and the Coulomb energy density
\citep{Chabanat1997}. The actual expressions can vary from one
parametrization to another.  For example, density-dependent
generalizations of the non-local ($t_1$ and $t_2$) terms of Skyrme
interactions have been added into the BSk functionals
\citep{Chamel2009}. In interactions developed for astrophysical
applications, the tensor coupling term is usually omitted
\citep{Goriely2010,Goriely2013,Goriely2016}. The KIDS interactions
have generalized the density dependent ($t_3$) term to several
$t_{3i}$ terms with corresponding exponents $\alpha_i$
\citep{Gil2019}. In our previous work \citep{Duan2024}, the
functionals Sky3 and Sky4 have been constructed combining the
modification of the density-dependent term with the density-dependent
generalizations of the $t_1$ and $t_2$ terms, namely, the functionals
contain not only the $t_{3i}$ terms but also the density-dependent
non-local $t_4$ and $t_5$ terms as in the extended BSk functionals.

As mentioned in Refs. \cite{Chamel2009,Duan2024}, the $(\nablav
\rho)^2$ terms cannot be transformed into $\rho \Delta \rho$ by
integration by parts in the extended Skyrme functionals, in contrast
to standard Skyrme functionals. Similarly, also the $(\nablav \otimes
\sv)^2=\sum_{ij}(\nabla_i s_j)^2$ terms cannot be transformed into
$\sv \cdot \Delta \sv$ by integration by parts
\cite{Pastore2015,Duan2024b}. Therefore, within the form chosen in
Ref. \cite{Duan2024}, the spin-dependent parts of the Skyrme energy
density functional can be written using standard notations as
\cite{Bender2003,Duan2023}
\begin{align}\label{eq:Skyrme energy density functional-ANM}
\varepsilon_{\text{spin}} =\; 
& C_0^{s} \sv^2 
+ C_1^{s} (\sv_n - \sv_p)^2  
+ C_0^{sT} ( \sv \cdot \vect{T} -\mathbbm{J}^2) 
\nonumber\\ &
+ C_1^{sT} [ (\sv_n - \sv_p) \cdot (\vect{T}_n - \vect{T}_p) - (\mathbbm{J}_n - \mathbbm{J}_p)^2] 
\nonumber\\ &
+ C_0^{\laplacian s} \sv \cdot \laplacian \sv 
+ C_1^{\laplacian s} (\sv_n - \sv_p) \cdot \laplacian (\sv_n - \sv_p)   
\nonumber\\&
+ C_0^{\nabla\otimes s} (\nablav\otimes \sv)^2
+ C_1^{\nabla\otimes s} [\nablav\otimes(\sv_n-\sv_p)]^2 
\nonumber\\ &
+ C_0^{\nabla J} ( \rho \nablav \cdot \vect{J} + \sv \cdot \nablav \times \vect{j}) 
\nonumber\\ &
+ C_1^{\nabla J} [(\rho_n - \rho_p) \nablav \cdot (\vect{J}_n - \vect{J}_p)
\nonumber\\ &
\phantom{{}+ C_1^{\nabla J} [} + (\sv_n - \sv_p) \cdot \nablav \!\times\! (\vect{j}_n - \vect{j}_p)]\,,
\end{align}
where (with the notation $y_i = t_i x_i$)
\begin{subequations}\label{eq:C_i-coefficients}
\begin{align}
C_0^{s} =\;&  {-\frac{1}{8}} ( t_0 -2y_0 ) - \sum_{i=1}^{3} \frac{1}{48} ( t_{3i} - 2y_{3i} ) \rho^{\alpha_i}, \\
C_1^{s} =\;&  {-\frac{1}{8}} t_0 -\sum_{i=1}^{3} \frac{1}{48} t_{3i} \rho^{\alpha_i}, \\
C_0^{sT}=\;& \frac{\eta_J}{16} [-(t_1-2y_1) + (t_2 +2y_2) \notag \\
&\phantom{ \frac{\eta_J}{16} [}-(t_4-2y_4) \rho^{\beta} + (t_5 +2y_5) \rho^{\gamma}], \\
C_1^{sT}=\;& \frac{\eta_J}{16}( -t_1 + t_2 - t_4 \rho^{\beta} + t_5 \rho^{\gamma}), \\
C_0^{\Delta s} =\;&  \frac{1}{32} [(t_1 -2y_1 )   + ( t_4 -2y_4 ) \rho^{\beta}], \\
C_1^{\Delta s} =\;&  \frac{1}{32}( t_1  +  t_4 \rho^{\beta} ), \\
C_0^{\nabla \otimes s} =\;&  {-\frac{1}{64}} [ (t_1-2y_1) + (t_2+2y_2) \notag \\
                &\phantom{{-\frac{1}{64}} [} + (t_4 - 2y_4) \rho^{\beta} +(t_5+2y_5) \rho^{\gamma}], \\
C_1^{\nabla \otimes s} =\;&  {-\frac{1}{64}} (t_1 + t_2 + t_4 \rho^{\beta} + t_5 \rho^{\gamma}), \\
C_0^{\nabla J} =\;&  {-\frac{3}{4}} W_0, \\
C_1^{\nabla J} =\;&  {-\frac{1}{4}} W_0.
\end{align}
\end{subequations}
The parameter $\eta_J$, in the notation of \cite{Bender2003},
indicates whether the contribution of the $t_1$ and $t_2$ terms to the
spin-orbit field should be included ($\eta_J=1$) or not
($\eta_J=0$). In our new functionals Sky3 and Sky4, we use $\eta_J=0$,
i.e., we have $C^{sT}_0=C^{sT}_1=0$ in these two functionals. For the
relation between the spin current $\mathbbm{J}$ and the vector spin
current $\vect{J}$ see e.g. Ref. \cite{Lesinski2007}.

Concerning the density dependence of the $C$ coefficients, in
Ref. \cite{Duan2024}, the choice was made to write them as polynomials
of $\rho^{1/3}$ (corresponding to an expansion in powers of
$k_F$). Hence, in Sky3 and Sky4, the exponents $\alpha_i$, $\beta$,
and $\gamma$ were not fitted. In the $t_3$ and $y_3$ terms, they are
$\alpha_i = i/3$, as in Ref. \cite{Gil2019}. Limiting the density
dependence of the $C^\rho$ coefficients (i.e., the $t_3$ and $y_3$
terms) to order $\rho$ ($i=3$) allowed us to precisely fit
microscopically computed PNM EoS from extremely low to the highest
densities. In order not to interfere with the EoS fit, the density
dependence of the effective mass term was limited to order
$\rho^{1/3}$ as explained in Ref. \cite{Duan2024}, and hence the
exponents are $\beta=\gamma=1/3$ in Sky3 and Sky4, while they have
different values in the extended Skyrme forces of the BSk family
\cite{Chamel2009}.

\begin{figure}
\begin{center}
\includegraphics[scale=0.5]{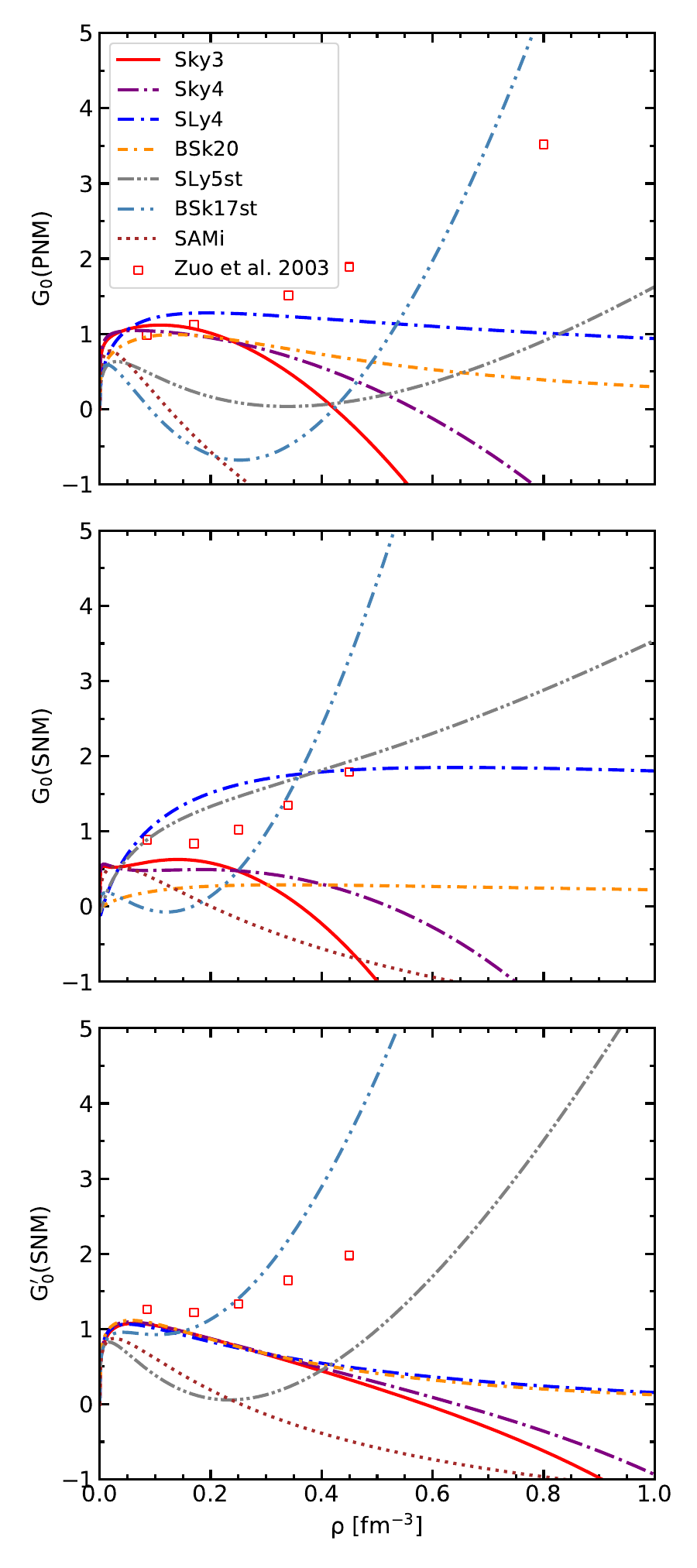} 
\caption{The Landau parameters $G_0$ in pure neutron matter (top
  panel) and $G_0$ and $G_0^{\prime}$ in symmetric nuclear matter
  (middle and bottom panels). The results are computed with Skyrme
  interactions Sky3, Sky4 \citep{Duan2024}, SLy4 \citep{Chabanat1998},
  BSk20 \citep{Goriely2010}, SLy5st \citep{Margueron2009}, BSk17st
  \citep{Margueron2009b}, and SAMi \cite{Roca-Maza2012}. The square
  symbols represent the microscopic BHF results of
  Refs. \cite{Zuo2003PNM,Zuo2003communications,Zuo2003SNM}.}
\label{fig:G-unconstrained}
\end{center}
\end{figure}
In terms of $C_0^s$, $C_1^s$, $C_0^{sT}$, and $C_1^{sT}$, the Landau
parameter $G_0$ in pure neutron matter can be written as
\begin{equation}\label{eq:G0-PNM}
G_0(\text{PNM}) = 2N_{0}[ C_0^s+C_1^s+k_F^2(C_0^{sT}+C_1^{sT})]\,,
\end{equation}
while the Landau parameters $G_0$ and $G_0^{\prime}$ in symmetric
nuclear matter can be written as
\begin{equation}\label{eq:G0-SNM}
G_0(\text{SNM}) = 4N_{0}(C_0^s+C_0^{sT}k_F^2),
\end{equation}
and
\begin{equation}\label{eq:G0'-SNM}
G_0^{\prime}(\text{SNM}) = 4N_{0}(C_1^s+C_1^{sT}k_F^2),
\end{equation}
where $N_0=\frac{m^*k_F}{\pi^2\hbar^2}$, and $k_F=(3\pi^2\rho)^{1/3}$
for pure neutron matter and $k_F=(\frac{3}{2}\pi^2\rho)^{1/3}$ for
symmetric nuclear matter.

The Landau parameters computed with Sky3 and Sky4 were already shown
in Ref. \cite{Duan2024}. We display them again in
Fig. \ref{fig:G-unconstrained} together with the results computed with
other Skyrme interactions, such as SLy4 \citep{Chabanat1998}, BSk20
\citep{Goriely2010}, SLy5st \citep{Margueron2009}, BSk17st
\citep{Margueron2009b}, and SAMi \cite{Roca-Maza2012}. SLy4 and BSk20
are, respectively, examples for standard and extended Skyrme
interactions. SLy5st and BSk17st represent Skyrme interactions whose
spin-dependent parts have been modified in an attempt to constrain
them using BHF results \citep{Margueron2009,Margueron2009b}. SAMi is a
standard Skyrme force but fitted with special attention to
Gamow-Teller resonances.  For comparison, the microscopic BHF results
\citep{Zuo2003PNM,Zuo2003SNM} \footnote{There is a typo in Table~1 of
  Ref. \cite{Zuo2003PNM} and $G_0$ in pure neutron matter for
  $\rho=0.8\fm^{-3}$ should read $3.52$ instead of $2.52$
  \cite{Zuo2025}.} are shown as square symbols. As we can see, the
results computed with Sky3 and Sky4 are close to the BHF values at
saturation density, but their evolutions with density are very
different from the BHF results. Other Skyrme interactions present not
only different evolutions but also different values at saturation
density. Of course, it is not a necessary condition that a reasonable
functional must reproduce the BHF results, that in addition do not
have a quantified uncertainty. But it is interesting that even in the
cases of SLy5st \citep{Margueron2009} and BSk17st
\citep{Margueron2009b}, which used the BHF results for the adjustment
of their additional parameters, there is not enough freedom to
reproduce the BHF results.

\begin{figure}
\begin{center}
\includegraphics[scale=0.5]{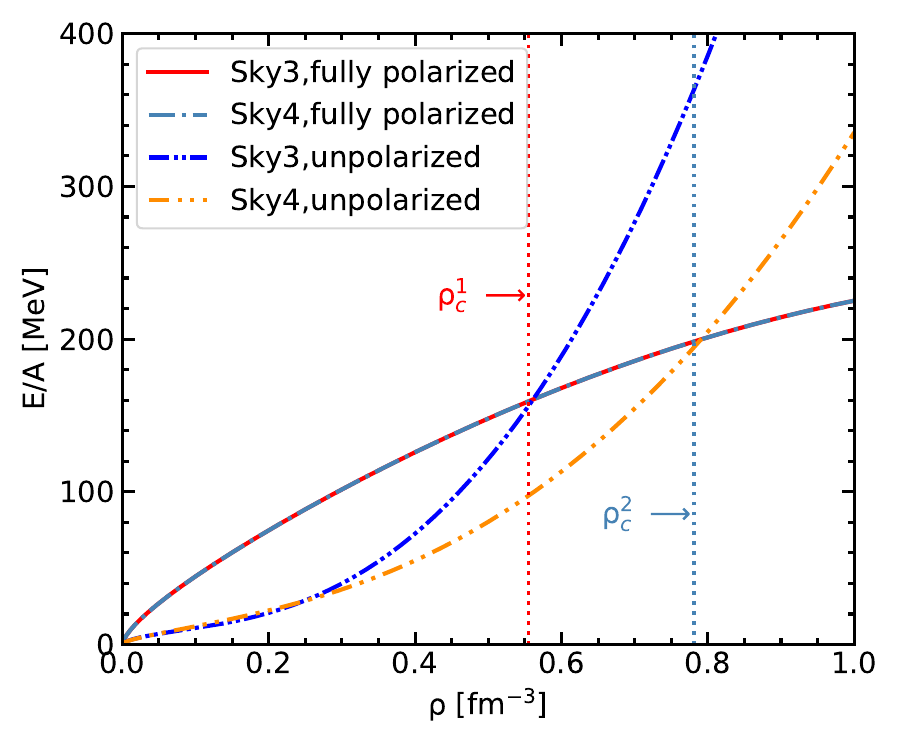} 
\caption{Equations of state of fully spin-polarized and unpolarized
  neutron matter computed with Sky3 and Sky4. The vertical red and
  steel-blue dotted lines represent the critical densities in pure
  neutron matter described using Sky3 and Sky4,
  respectively.}\label{fig:EoS-unconstrained}
\end{center}
\end{figure}

In the case of Sky3, Sky4, and SAMi, the situation is worse because
above some critical densities they give $G_0 < -1$, violating the
Landau stability criterion. To further illustrate this instability
problem, we show in Fig. \ref{fig:EoS-unconstrained} the equations of
state (EoS) of fully spin-polarized and unpolarized neutron matter
computed with Sky3 and Sky4. For both Sky3 and Sky4, the fully
polarized neutron matter becomes energetically favorable over the
unpolarized one at densities above the respective critical densities
indicated by the red and steel-blue dotted lines. The fundamental
cause of this behaviour is the wrong values of the Landau parameters
mentioned above. Otherwise, one would get a stiffer EoS for fully
polarized than for unpolarized neutron matter. Therefore, the solution
is to revise Sky3 and Sky4 to obtain realistic values of the Landau
parameters.

\section{Parameters and nuclear-matter properties of the new Skyrme
  functionals}
\label{sec:constraints to the spin dependent terms}
\subsection{Determination of the spin-dependent terms of the new Skyrme
  functionals}
\label{subsec:strategies}
From now on, we treat the functional as a general density functional
instead of a functional derived from a density-dependent two-body
force. According to Ref. \cite{Bender2002}, the spin-independent and
spin-dependent terms in the energy functional can be adjusted
independently. Actually, the common choice $\eta_J=0$ (which we also
adopted in \cite{Duan2024} for Sky3 and Sky4) is already an example
for this philosophy since the functional derived from the two-body
force would correspond to $\eta_J=1$.

To distinguish them from the previous functionals Sky3 and Sky4, we
will denote the new functionals with independently adjusted
spin-dependent terms by Sky3s and Sky4s, respectively. First of all,
analogously to the common choice $\eta_J=0$, we set
\begin{equation}\label{eq:Cnablas-Cdeltas}
C_0^{\nabla \otimes s}=C_1^{\nabla \otimes s} = C_0^{\Delta s} = C_1^{\Delta s} =0,
\end{equation}
since these coefficients are completely unconstrained by the
observables (energies and radii of doubly-magic nuclei) to which we
have fitted the parameters.

\begin{figure}
\begin{center}
\includegraphics[scale=0.5]{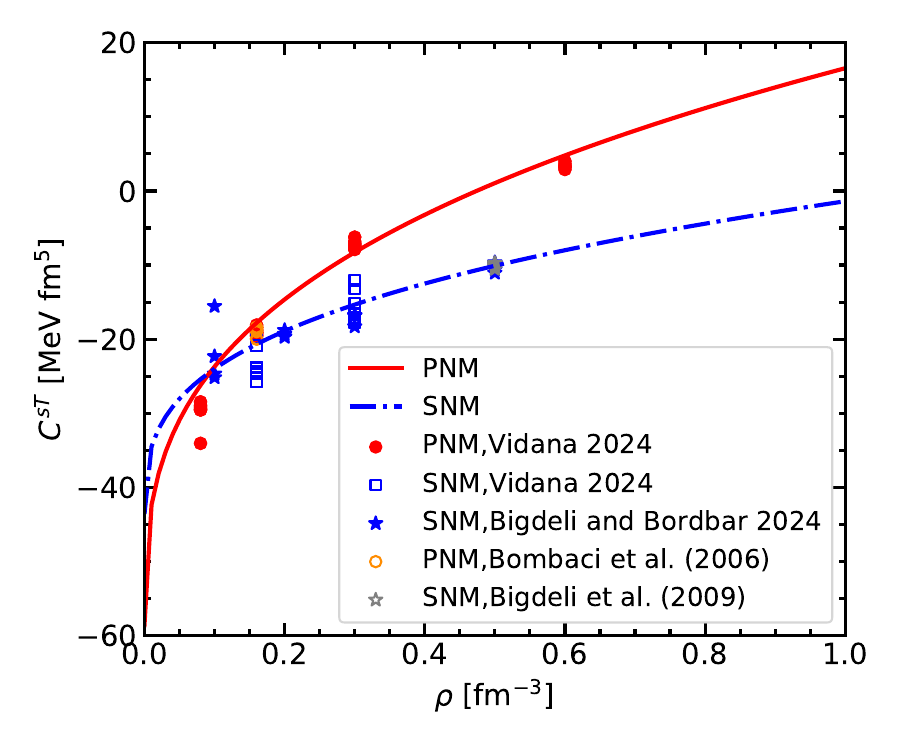} 
\caption{Fits to microscopic results of
  \cite{Bombaci2006,Bigdeli2009,Bigdeli2024,Vidana2024} for $C^{sT}$
  as functions of density in pure neutron matter (red) and symmetric
  nuclear matter (blue). The scattering of the microscopic results for
  fixed densities comes from deducing $C^{sT}$ at different values of
  the polarization.\label{fig:CsT}}
\end{center}
\end{figure}
We now focus on constraining the remaining coefficients $C_0^s$,
$C_1^s$, $C_0^{sT}$, and $C_1^{sT}$. As mentioned before, the choice
$\eta_J=0$ implies that, in Sky3 and Sky4, we have $C_0^{sT} =
C_1^{sT} = 0$. This means that in spin-polarized matter, the spin up
and down nucleons have the same effective mass, since
\begin{equation}
  \frac{m}{m^*_\uparrow}-\frac{m}{m^*_\downarrow}
  = \frac{4m}{\hbar^2}C^{sT}(\rho_\uparrow-\rho_\downarrow)\,,
\end{equation}
where $C^{sT} = C_0^{sT}$ in the case of symmetric nuclear matter and
$C^{sT} = C_0^{sT}+C_1^{sT}$ in the case of pure neutron
matter. However, the BHF calculations of Ref. \cite{Bombaci2006} for
spin-polarized neutron matter and calculations within the lowest-order
constrained variational (LOCV) method for spin-polarized symmetric
matter \cite{Bigdeli2009} found a splitting of the effective masses of
spin up and down nucleons.

Since the spin-independent and spin-dependent terms in the energy
functional can be adjusted independently \citep{Bender2002}, we will
determine the $C^{sT}$ terms from these splittings rather than
relating them to the parameters $t_1\dots y_5$ that were adjusted only
to spin-independent observables.  Assuming the same form of the
density dependence as in Eq.~\eqref{eq:C_i-coefficients} with
$\beta=\gamma=\frac{1}{3}$ as in Sky3 and Sky4, we write
$C^{sT}_{\iso}$ ($\iso=0,1$) as
\begin{equation}\label{eq:CsT-rewrite}
C^{sT}_{\iso}=C^{sT}_{\iso 0} +C^{sT}_{\iso 1} \rho^{\frac{1}{3}}.
\end{equation}
Then $C^{sT}_{00}$ and $C^{sT}_{01}$ can be obtained by fitting the
splitting between the up and down nucleon effective masses for
spin-polarized symmetric nuclear matter, while
$C^{sT}_{00}+C^{sT}_{10}$ and $C^{sT}_{01}+C^{sT}_{11}$ can be
obtained by fitting the splitting between the up and down neutron
effective masses for spin-polarized neutron matter. To get information
about the density dependence, the results published in
Refs. \cite{Bombaci2006,Bigdeli2009} are not enough and we obtained
more data at different densities directly from the authors of these
papers \cite{Bigdeli2024,Vidana2024}.

The results computed by Ref. \cite{Bigdeli2024} used only the Argonne
V18 two-body force, while Ref. \cite{Vidana2024} also added the effect
of the phenomenological three-body force (TBF) Urbana IX (reduced to a
density dependent two-body force). Using these results is slightly
inconsistent because the Sky3 and Sky4 effective masses were adjusted
to the results of Ref. \cite{Baldo2014} obtained with a
microscopically derived TBF~\cite{Grange1989}, which was also employed
in the calculation of the Landau parameters
\cite{Zuo2003PNM,Zuo2003communications,Zuo2003SNM}.  While the TBF has
almost no effect at low density, it causes large uncertainties at high
density. However, we have no choice because to our knowledge there are
no results for the splitting of up and down effective masses computed
with AV18 and the microscopic TBF in the present literature, and
recomputing it using a microscopic TBF is a difficult task that
clearly goes beyond the scope of this work. Moreover, our purpose is
to get realistic spin properties instead of reproducing all nuclear
matter properties computed using the microscopic TBF. Therefore, this
imperfection should be acceptable for now.

In Fig. \ref{fig:CsT}, we show $C^{sT}$ as functions of density in
pure neutron matter and symmetric nuclear matter. As we can see,
$C^{sT}_{\iso}$ cannot be considered constant. Obviously, the given
fit is not the only possible one, but since we are fitting our
parameters to theoretical results whose uncertainties are unknown, it
is difficult to give a quantitative estimate of the uncertainties of
the fitted $C^{sT}$ coefficients.

As mentioned above, the $C_{\iso}^s$ and $C_{\iso}^{sT}$ coefficients
can be related to the Landau parameters $G_0$ and $G_0^{\prime}$ in
pure neutron matter and symmetric nuclear matter. Therefore, the next
step is to obtain realistic values of the Landau parameters for our
two functionals. As we did above for the $C^{sT}$ coefficients and in
Ref. \cite{Duan2024} for the effective masses, we will again use the
results of microscopic BHF calculations for this purpose.

We now rewrite $C_0^s$ and $C_1^s$ as
\begin{equation}\label{eq:re-C0s}
  C_0^s = -\frac{1}{8} (t_0-2y_0)
  - \sum_{i=1}^{3} \frac{1}{48} (t_{3i}^{s} - 2y_{3i}^{s}) \rho^{\alpha_i},
\end{equation}
and 
\begin{equation}\label{eq:re-C1s}
C_1^s=-\frac{1}{8} t_0 - \sum_{i=1}^{3} \frac{1}{48} t_{3i}^{s} \rho^{\alpha_i}.
\end{equation}
This means $t_{3i}^s$ and $y_{3i}^s$ are different from $t_{3i}$ and
$y_{3i}$ in the spin-independent terms, but $t_0$ and $y_0$ are kept
because the interaction is dominated by two-body interaction at very
low densities. The $C^s_{\iso}$ coefficients ($\iso=0,1)$ can also be
expressed as
\begin{equation}\label{eq:Cst} 
  C^s_{\iso}= C^s_{\iso 0} + C^s_{\iso 1} \rho^{\frac{1}{3}}
  + C^s_{\iso 2} \rho^{\frac{2}{3}} + C^s_{\iso 3} \rho,
\end{equation}
with
\begin{subequations}\label{eq:CsTi}
\begin{align}
C^s_{00} & = -\frac{1}{8} (t_0-2y_0), &\\
C^s_{0i} & = -\frac{1}{48} (t_{3i}^{s} - 2y_{3i}^s) &(i=1,2,3),\\
C^s_{10} & = -\frac{1}{8} t_0, & \\
C^s_{1i} & = -\frac{1}{48} t_{3i}^{s} &(i=1,2,3).
\end{align}
\end{subequations}
Hence, only $C^s_{00}$ and $C^s_{10}$ are defined by the values of
$t_0$ and $y_0$ of Sky3 or Sky4, respectively. We determine $C^s_{\iso
  1}\dots C^s_{\iso 3}$ by fitting only the Landau parameters $G_0$
and $G_0^{\prime}$ in symmetric nuclear matter to those computed using
the microscopic BHF theory in Ref. \cite{Zuo2003SNM}, because we found
a problem while analyzing the results for pure neutron matter in
Refs. \cite{Zuo2003PNM,Zuo2003communications} (the results presented
in these two references are the same), as described below. According
to Ref. \cite{Zuo2003communications}, $G_0$ in pure neutron matter is
computed from the spin susceptibility $\chi$ as
\begin{equation}\label{eq:G0-chi}
G_0= \frac{m^*}{m} \left(\frac{\chi}{\chi^{}_F}\right)^{-1} -1,
\end{equation}
where $\chi^{}_F$ is the spin susceptibility of an ideal gas. By
analyzing the different results for $G_0$ and $\chi$ shown in
Refs. \cite{Zuo2003PNM,Zuo2003communications}, one finds that in the
calculation with TBF, the values of the factor $\frac{m^*}{m}$ are
almost the same as those computed without TBF. It is worth noting that
Refs. \cite{Zuo2003PNM,Zuo2003communications} give almost the same
$\frac{m^*}{m}$ as Ref. \cite{Baldo2014} in the two-body force case,
while they give very different results when including the TBF, since
the TBF seems to have almost no effect on $\frac{m^*}{m}$ in
Refs. \cite{Zuo2003PNM,Zuo2003communications} while it has a strong
effect in Ref. \cite{Baldo2014}. Since the Sky3 and Sky4 effective
masses in PNM are similar to the results from Ref. \cite{Baldo2014}
(to which they were adjusted) instead of those from
Refs. \cite{Zuo2003PNM,Zuo2003communications}, we should not fit $G_0$
of Refs. \cite{Zuo2003PNM,Zuo2003communications} directly, because it
is normalized with $N_0=\frac{m^*k_F}{\pi^2 \hbar^2}$ that contains
the neutron effective mass. However, we can fit
$\tfrac{\chi}{\chi^{}_F}$ in essence. Therefore, we recompute $G_0$
from Eq. \eqref{eq:G0-chi} using the results for
$\tfrac{\chi}{\chi^{}_F}$ given in
Refs. \cite{Zuo2003PNM,Zuo2003communications} and for $\tfrac{m^*}{m}$
from Ref. \cite{Baldo2014}. Let us mention that in the case of SNM,
this problem does not exist, because the ratios $\frac{m^*}{m}$ that
can be deduced from the SNM results given in Ref. \cite{Zuo2003SNM}
agree well with those of Sky3 and Sky4. The results of the fits are
listed in Table \ref{table:t3i-y3i-refitting}.

Again, as in the case of the $C^{sT}$ coefficients, since we are
fitting to theoretical calculations with unknown uncertainties, we
cannot provide errors estimates for the fitted parameters. Just from
comparing the results for the Landau parameters obtained in
Refs. \cite{Zuo2003PNM,Zuo2003communications,Zuo2003SNM} with and
without TBF and considering the large uncertainties of the TBF, it is
clear that there may be substantial errors in the Landau parameters.

\begin{table}
\centering
\caption{Values of the new parameters $C^{sT}_{\iso i}$, $C^{s}_{\iso
    i}$, and $W_0$ for Sky3s and Sky4s. All other parameters are the
  same as in Sky3 and Sky4, respectively.}
\label{table:t3i-y3i-refitting}
\begin{ruledtabular}
\begin{tabular}{lcc}

&Sky3s&Sky4s\\
\hline
$C^{sT}_{00}$ (MeV fm$^{5}$) & -43.6 & -43.6 \\
$C^{sT}_{01}$ (MeV fm$^{6}$) & 42.2 & 42.2 \\
$C^{sT}_{10}$ (MeV fm$^{5}$) & -15.0  & -15.0 \\
$C^{sT}_{11}$ (MeV fm$^{6}$) & 33.0  & 33.0 \\
$C^s_{00}$ (MeV fm$^{3}$) & 338.7 & 314.3 \\
$C^s_{01}$ (MeV fm$^{4}$) & -1059.4 & -945.7 \\
$C^s_{02}$ (MeV fm$^{5}$) & 1414.6 & 1239.9 \\
$C^s_{03}$ (MeV fm$^{6}$) & -406.7 & -318.0 \\
$C^s_{10}$ (MeV fm$^{3}$) & 225.1 & 242.9 \\
$C^s_{11}$ (MeV fm$^{4}$) & -385.3 & -468.3 \\
$C^s_{12}$ (MeV fm$^{5}$) & 336.7 & 464.2 \\
$C^s_{13}$ (MeV fm$^{6}$) & -32.7 & -97.4 \\
$W_0$ (MeV fm$^{5}$) & 117.4 & 114.1
\end{tabular}
\end{ruledtabular}
\end{table}

Since we have now non-vanishing $C_{\iso}^{sT}$, the spin-orbit
splitting in finite nuclei is modified, because the spin-orbit field
$U_{ls,q}$ (in the notation of \cite{Reinhard1991}, $q = n,p$)
acquires an additional term
\begin{equation}
U_{ls,q}^{\text{(add)}} = (C^{sT}_1-C^{sT}_0)J - 2C^{sT}_1 J_q\,.
\end{equation}
Because of the density dependence of $C^{sT}_\iso$, there are also
additional terms in the mean-field potential,
\begin{equation}
  U^{\text{(add)}} = \frac{1}{2}(C_1^{sT\,\prime}-C_0^{sT\,\prime})J^2
  -C_1^{sT\,\prime} (J_n^2+J_p^2)\,,
\end{equation}
(with $C_{\iso}^{sT\,\prime} = dC_{\iso}^{sT}/d\rho = C_{\iso 1}^{sT}
\rho^{-2/3}/3$), and in the rearrangement energy,
\begin{equation}
    E_{\text{rearr}}^{\text{(add)}} = -\int\! d^3r\, \frac{1}{2}\,\rho\,U^{\text{add}}.
\end{equation}

In order to still get good binding energies and radii, we therefore
have to refit the spin-orbit parameter $W_0$. Refitting only $W_0$ and
keeping the other parameters as in Ref. \cite{Duan2024} is sufficient
because after the refitting, the $\chi^2$ of binding energies and
radii as defined in \cite{Duan2024} is 4.48 for Sky3s and 5.55 for
Sky4s, which is almost as good as the $\chi^2$ of 4.08 and 5.07
obtained for Sky3 and Sky4. The new values of $W_0$ can also be found
in Table \ref{table:t3i-y3i-refitting}. We tried refitting also the
other parameters, but the resulting improvement of $\chi^2$ was
negligible.

\subsection{The lowest-order Landau parameters in nuclear matter and
  spin-polarized neutron matter EoS}
\label{subsec:lowest-order Landau parameters}
\begin{figure}
\begin{center}
\includegraphics[scale=0.5]{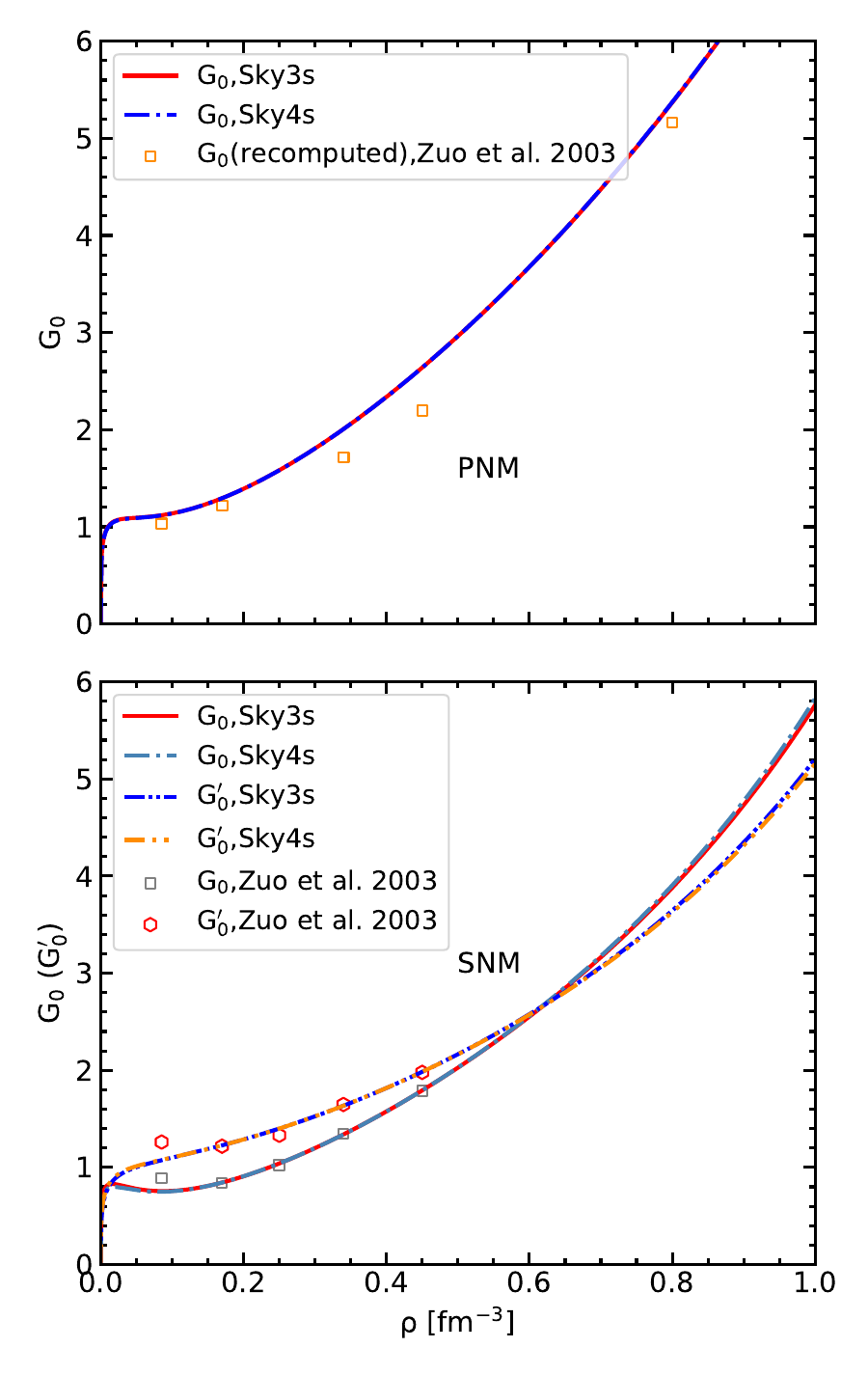}
\caption{The Landau parameters $G_0$ in pure neutron matter (upper
  panel) and $G_0$ and $G_0^{\prime}$ in symmetric nuclear matter
  (lower panel) computed with Sky3s and Sky4s. The $G_0$ in pure
  neutron matter recomputed from the results of
  Refs. \cite{Zuo2003PNM,Zuo2003communications} as explained in
  Sec. \ref{subsec:strategies}, and the results for symmetric nuclear
  matter from Ref. \cite{Zuo2003SNM} are also
  shown.  \label{fig:Landau-parameters-PNM-SNM}}
\end{center}
\end{figure}

\begin{figure}
\begin{center}
\includegraphics[scale=0.5]{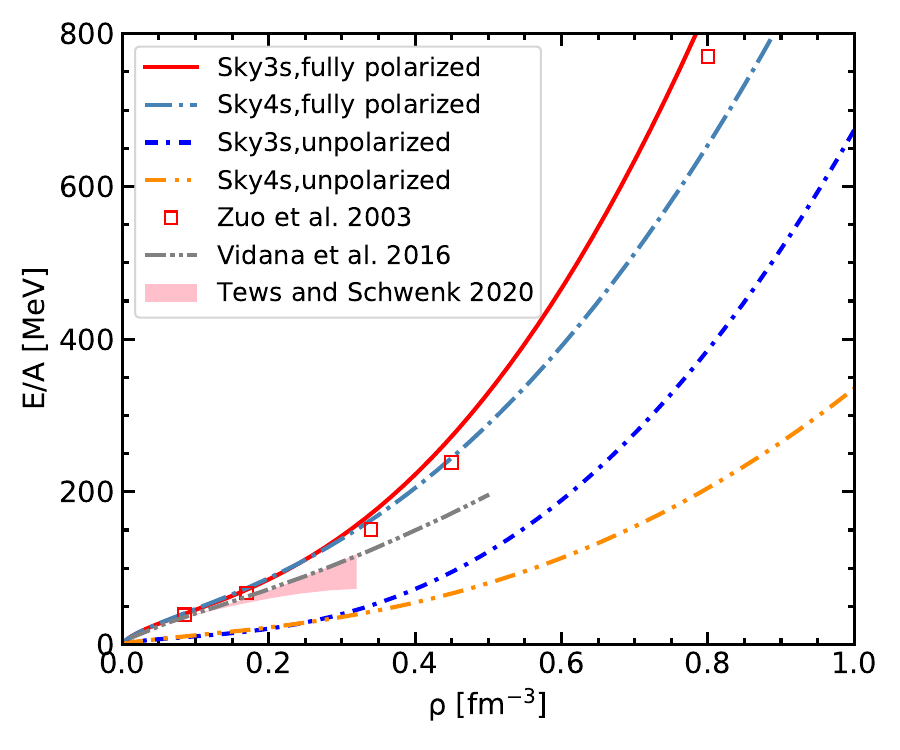} 
\caption{EoS in fully spin-polarized and unpolarized neutron matter
  for Sky3s and Sky4s. The results from
  Refs. \cite{Zuo2003PNM,Vidana2016,Tews2020} for polarized neutron
  matter are also shown for
  comparison.}\label{fig:EoS-PNM-polarization-nopolarization}
\end{center}
\end{figure}
In this section, we present some nuclear matter properties described
by the two new interactions with the new parameters as explained
above.

In Fig. \ref{fig:Landau-parameters-PNM-SNM}, we show the Landau
parameters $G_0$ in pure neutron matter (upper panel) and $G_0$ and
$G_0^{\prime}$ in symmetric nuclear matter (lower panel). For
comparison, we also present $G_0$ in pure neutron matter recomputed
from the results of Refs. \cite{Zuo2003PNM,Zuo2003communications} as
described in Sec. \ref{subsec:strategies}, and $G_0$ and
$G_0^{\prime}$ in symmetric nuclear matter from
Ref. \cite{Zuo2003SNM}. All values are larger than 0 ($> -1$). This
means that there is no ferromagnetic instability in spin-1
channels. The $G_0$ and $G_0^{\prime}$ in symmetric nuclear matter at
densities $0.17-0.45\fm^{-3}$ agree within 10\% with the BHF results
(only for the points at the lowest density, $0.08\fm^{-3}$, the
deviations are slightly larger, i.e., within approximately $15\%$). At
low densities, the agreement in SNM is not as good as in PNM but still
satisfactory. Besides, $G_0^{\prime}\approx 1.20$ for Sky3s and
$G_0^{\prime} \approx 1.19$ Sky4s at saturation density. These values
also agree well with the result $G_0^{\prime} \approx1.2$ of
Ref. \cite{Bender2002} obtained from Gamow-Teller resonances in finite
nuclei.

In Fig. \ref{fig:EoS-PNM-polarization-nopolarization}, we show the EoS
of fully spin-polarized and unpolarized neutron matter for Sky3s and
Sky4s. The microscopic BHF results \citep{Zuo2003PNM,Vidana2016} and
the AFDMC calculation with the N$^2$LO $V_{E1}$ of
Ref. \cite{Tews2020} are also shown for comparison. The difference
between the two microscopic BHF results is that they choose different
three-body forces. In Ref. \cite{Zuo2003PNM}, the microscopic TBF is
chosen, while the phenomenological TBF is chosen in
Ref. \cite{Vidana2016}. From the comparison between the EoS of
polarized and unpolarized neutron matter, we can see that
spin-polarized neutron matter always has higher energy than
unpolarized one at all densities for both Sky3s and Sky4s. This is
only possible because now also the energy of polarized matter
increases strongly at high density. This agrees qualitatively with the
results in Refs. \cite{Zuo2003PNM,Vidana2016} Not surprisingly, our
results have a better agreement with those of Ref. \cite{Zuo2003PNM}
(red squares in Fig. \ref{fig:EoS-PNM-polarization-nopolarization}),
because our spin Landau parameters are determined using the
microscopic results from that calculation, and it shows that our way
to recompute $G_0$ from $\chi/\chi_F$ and $m^*/m$ is reasonable. Of
course, the agreement cannot be perfect, because the spin Landau
parameters determine only the difference between polarized and
unpolarized EoS, and the unpolarized EoS of Sky3 and Sky4 are
different from the one of Ref. \cite{Zuo2003PNM}.

We can see that Sky3s and Sky4s give a slightly stiffer EoS in
spin-polarized neutron matter than BHF results with phenomenological
TBF \cite{Vidana2016} (grey curve) and AFDMC calculations
\cite{Tews2020} (light red band), even at densities below
$2\rho_0$. We expect that our results would agree better with those of
Ref. \cite{Vidana2016} if we had fitted the spin Landau parameters
computed with the phenomenological TBF, but this would not have been
consistent with our effective masses. All the above is also evidence
that there is an uncertainty coming from the TBF. Nevertheless, in
particular in the context of neutrino rates, the Landau parameters are
more relevant than the EoS of spin-polarized matter, and we believe
that our new interactions will be useful in spite of these
uncertainties.

\subsection{Instabilities of neutron star matter described by the two new
  interactions}
\label{subsec:instabilities}
There can be singularities at zero energy transfer in response
functions of nuclear matter. They indicate the appearance of
instabilities. The positions of instabilities should be determined to
see whether the new interactions can describe neutron star matter. We
will compute and show the results for $\beta$-stable neutron star
matter in this section.

In the channel $S=0$ ($S$ is the spin), the so-called spinodal
instability \citep{Ducoin2008} is physical. This spinodal instability
represents the liquid-gas phase transition, which has been observed in
symmetric nuclear matter in experimental studies
\citep{Suraud1989}. In the case of neutron star matter, it corresponds
to the core-crust transition. This transition can strongly affect the
transport process of neutrinos in hot and dense nuclear matter
\citep{Margueron2009,Reddy1999,Navarro1999}. In other words, this
transition can affect the supernova explosion and proto-neutron star
evolution because neutrinos play a crucial role in these astrophysical
processes.

However, other instabilities are unphysical. For small momentum
transfer $q\approx0$, we can also use the Landau stability criterion
as in the cases of pure neutron matter and symmetric nuclear
matter. In general, except for the spinodal instability, the
instabilities of neutron star matter (zero temperature) occur at
densities that are near the critical densities of pure neutron matter
and symmetric nuclear matter \citep{Davesne2014,Pastore2015}.

For different values of the baryon number density $\rho$ of
neutron-star matter and of the momentum transfer $q$, we can find the
onset of the instability from the condition \citep{Pastore2014}
\begin{align}\label{eq:solve-rhoc}
\frac{1}{\Pi^{(S,M,\iso)}_{\text{RPA}} (q,\omega=0)} = 0,
\end{align}
($M$ is the projection of spin along the direction of $\qv$, $\iso$ is
the isospin) to detect poles in the RPA response function
$\Pi_{\text{RPA}}$ after determining the proton fraction $Y_p$ through
the $\beta$-equilibrium condition at zero temperature.

\begin{figure}
\begin{center}
\includegraphics[scale=0.5]{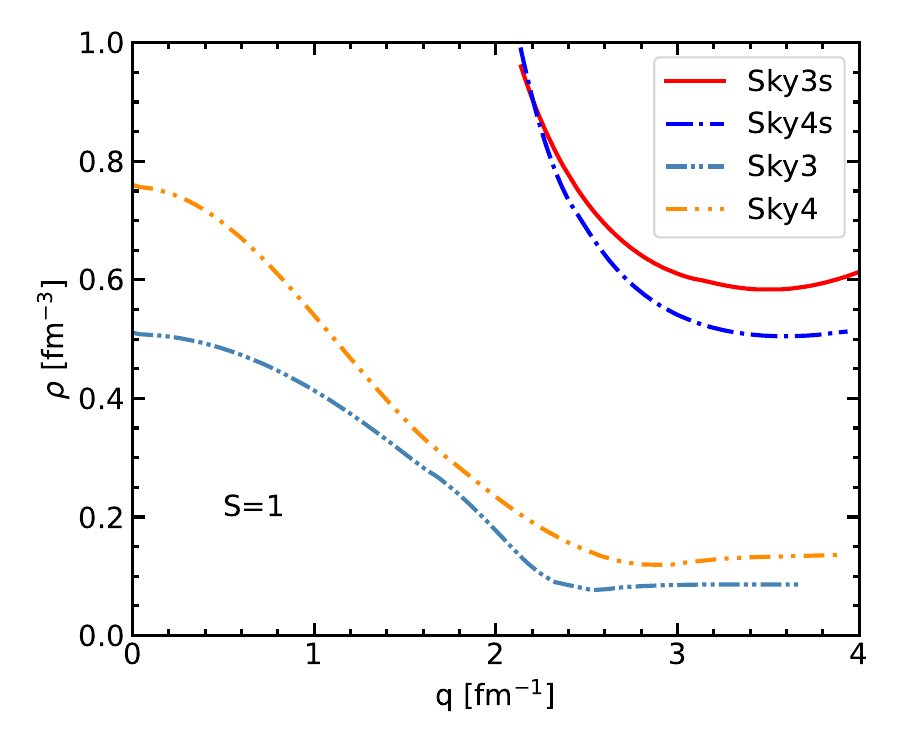} 
\caption{The onset of spin-1 instabilities in the $q-\rho$ plane for
  Sky3s and Sky4s for neutron star matter in $\beta$ equilibrium,
  compared to those of the previous unconstrained parametrizations
  Sky3 and Sky4.} \label{fig:rho-q-T0}
\end{center}
\end{figure}

We show the positions of the onset of the instabilities for Sky3s and
Sky4s in the $q-\rho$ plane in Fig. \ref{fig:rho-q-T0}. The results
for Sky3 and Sky4 are also shown for comparison. We show only the
results of spin-1 channels since the results of spin-0 channels have
been shown in Ref. \cite{Duan2024}.

In the case of the unconstrained interactions Sky3 and Sky4, we can
see from Fig. \ref{fig:rho-q-T0} that the instabilities can be found
at densities that are lower than the maximum central density of a
neutron star ($\rho_{\text{max}}=0.9248\fm^{-3}$ for Sky3,
$\rho_{\text{max}}=1.1264\fm^{-3}$ for Sky4 \citep{Duan2024}) even at
lower momentum transfers. Also, at higher momentum transfers, Sky3 and
Sky4 predict instabilities of neutron star matter even when the baryon
number density is lower than saturation density $\rho_0$. This is not
acceptable.

In the case of the new constrained interactions Sky3s and Sky4s, we
can see from Fig. \ref{fig:rho-q-T0} that the unphysical ferromagnetic
instabilities in spin-1 channels can be found in the full RPA
responses of neutron star matter only at high densities and high
momentum transfers. At densities near the maximum central densities of
neutron stars predicted by Sky3s and Sky4s, there is no unphysical
ferromagnetic instability until the momentum transfer is higher than
about $2\fm^{-1}$. At such high momentum transfers, Skyrme
interactions should not be used. But these momentum transfers do not
appear in neutrino scattering.

\section{Response functions and neutrino scattering rates in neutron-star
  matter}
\label{sec:responses-neutrino-rates}
\begin{figure}
\begin{center}
\includegraphics[scale=0.5]{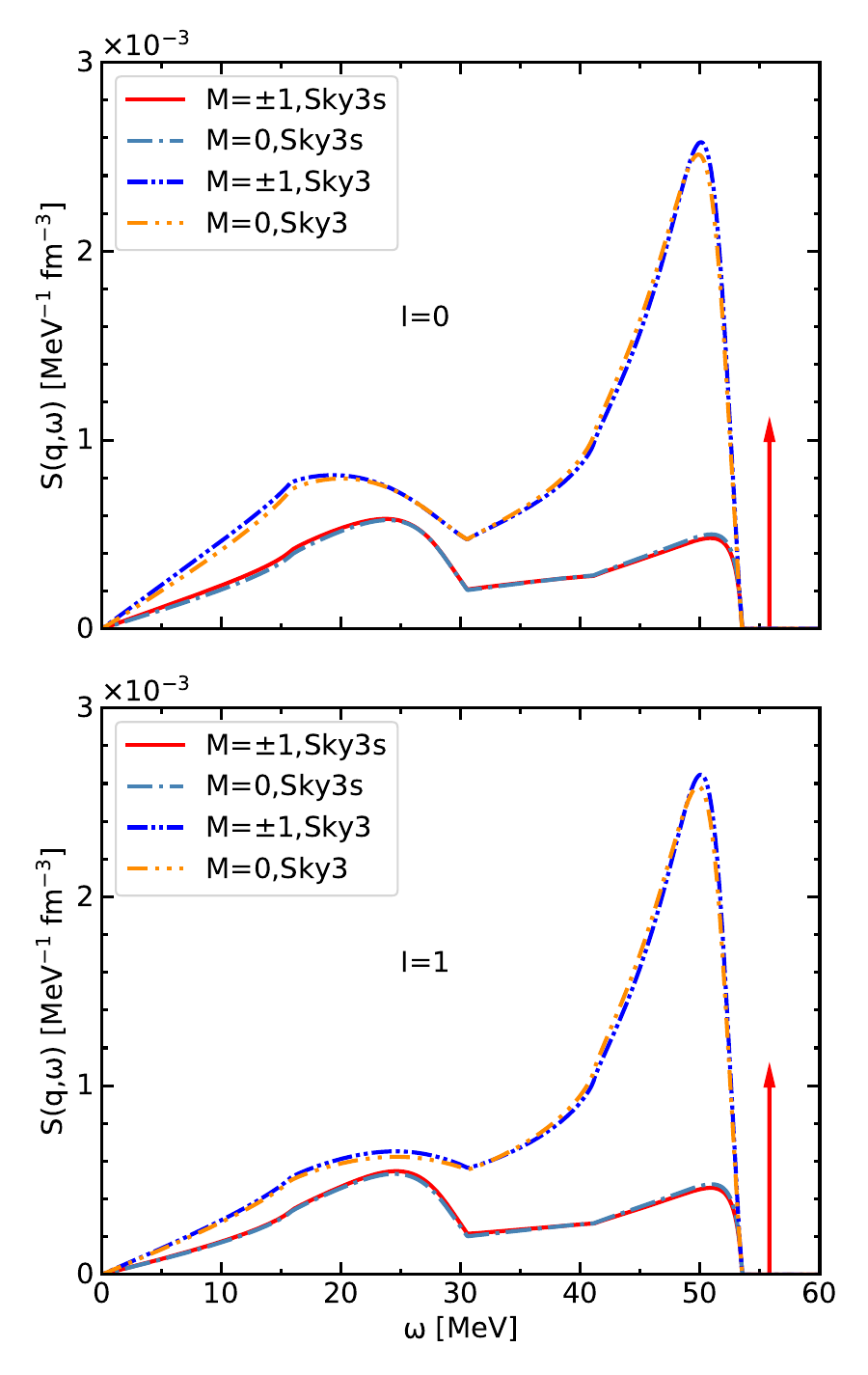}
\caption{Spin-1 channel response functions computed with
  Sky3s. Neutron star matter is at zero temperature and $\beta$
  equilibrium. $\rho=0.25\fm^{-3}$, $Y_p=0.0646$, $q=0.5\fm^{-1}$. The
  results computed with Sky3 are also shown for comparison. Upper
  panel: isoscalar response functions ($\iso=0$); lower panel:
  isovector response functions
  ($\iso=1$).}\label{fig:response-functions-spin1}
\end{center}
\end{figure}

\begin{figure}
\begin{center}
\includegraphics[scale=0.5]{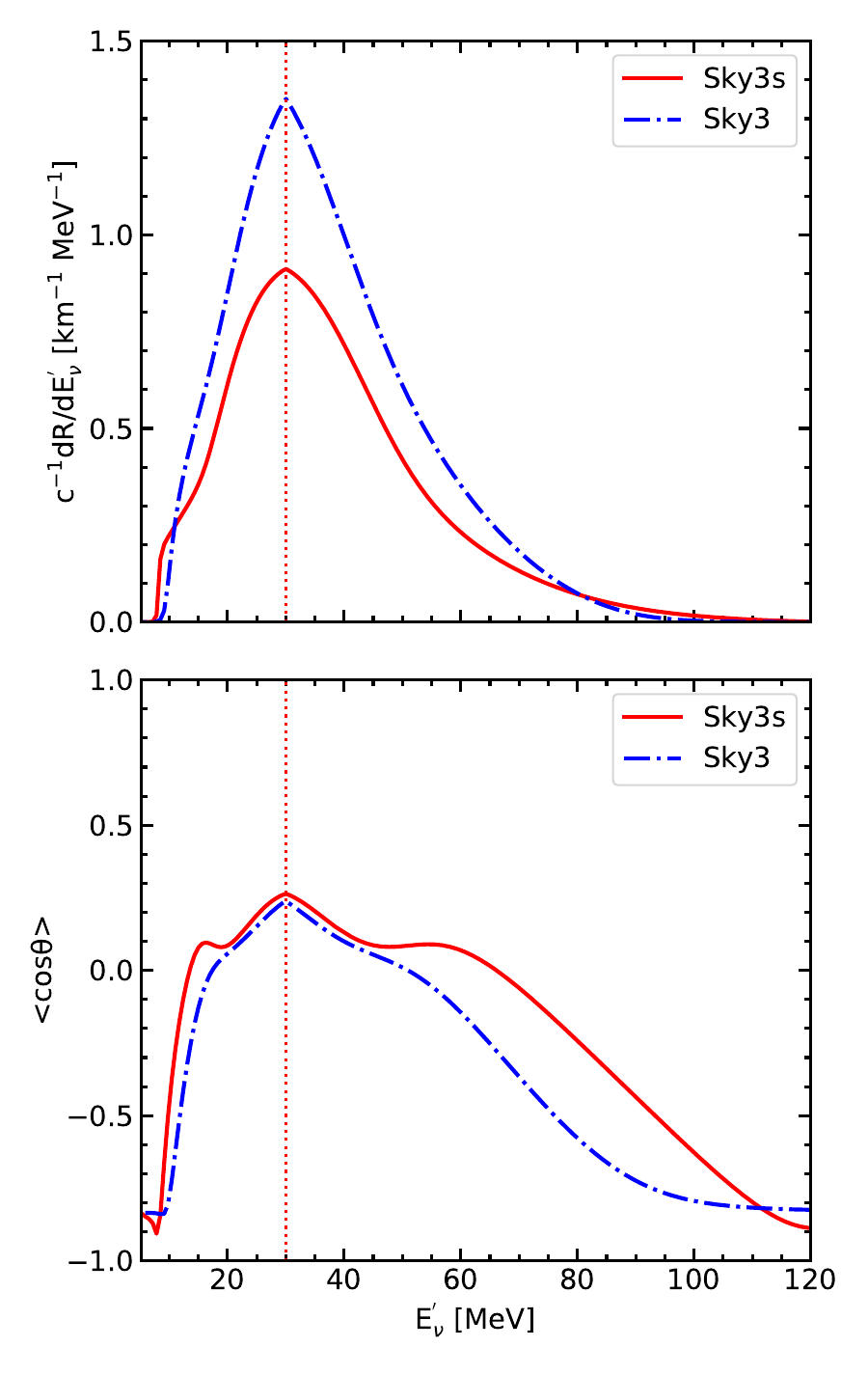} 
\caption{Differential neutrino scattering rates as a function of the
  final neutrino energy $E_{\nu}^{\prime}$ with the fixed initial
  neutrino energy $E_{\nu}=3T$ for Sky3s (upper panel) and the
  corresponding evolution of the average cosine of the scattering
  angle (lower panel). $T=10\MeV$, $\rho=0.25\fm^{-3}$,
  $Y_p=0.0646$. The red dotted lines correspond to the initial
  neutrino energy. The results computed with Sky3 are also shown for
  comparison.}\label{fig:neutrino-rates-comparison}
\end{center}
\end{figure}
In this section, we show examples for full RPA response functions for
$S=1$ channels and neutrino scattering rates computed using Sky3s.

Pastore et al. presented a detailed review of response functions
computed using Skyrme energy density functional
\citep{Pastore2015}. Another equivalent method to compute response
functions was proposed in Ref. \cite{Urban2020}, which has been
generalized for use in asymmetric nuclear matter and finite energy
transfer \citep{Duan2023}. Here, we adopt the method described in
Ref. \cite{Duan2023} to compute the RPA response functions of neutron
star matter. Following the steps explained in Section
\uppercase\expandafter{\romannumeral2} B of Ref. \cite{Duan2023}, the
expressions of RPA response functions in asymmetric nuclear matter can
be obtained.

In general, the dynamical structure factors can be related to the RPA
response functions as
\begin{align}\label{eq:relation-S-Pi}
  S^{(S,M,\iso)}(\qv,\omega) = -\frac{1}{\pi} \frac{1}{1-e^{-\omega/T}}
  \Im \Pi^{(S,M,\iso)}_{\text{RPA}}(\qv,\omega).
\end{align}

We show the spin-1 dynamical structure factors $S(q,\omega)$ for Sky3s
in neutron star matter at zero temperature under the condition of
$\beta$ equilibrium in Fig. \ref{fig:response-functions-spin1}. The
upper and lower panels show the isoscalar ($\iso=0$) and isovector
($\iso=1$) responses, respectively, for baryon number density
$\rho=0.25\fm^{-3}$, proton fraction $Y_p=0.0646$, and momentum
transfer $q=0.5\fm^{-1}$. The red arrows represent the positions of
the zero-sound modes for Sky3s. The results computed with Sky3 are
also shown for comparison. As we can see, the responses computed with
Sky3 are larger than those computed with Sky3s. Besides, the
zero-sound modes lie in the particle-hole continuum for Sky3. These
differences are due to the fact that the unconstrained Landau
parameters of Sky3 are not repulsive enough, which can affect
computations of neutrino rates in (proto-)neutron star and supernova
matter.

As mentioned in the introduction, neutrinos play a crucial role in
explaining some astrophysical phenomena. The interactions between
neutrinos and nuclear matter should not be ignored. In particular,
neutrino scattering is as important as neutrino absorption.

We compute the differential neutrino scattering rate and average
cosine of the scattering angle using the method described in Section
\uppercase\expandafter{\romannumeral4} C of Ref. \cite{Duan2023} and
show the results in Fig. \ref{fig:neutrino-rates-comparison}.  To
include the contribution of the zero-sound modes, we compute at
temperature $T=10\MeV$ (so that the zero-sound mode gets a finite
width) but keep the baryon number density and proton fraction as in
Fig. \ref{fig:response-functions-spin1}.  The results are plotted as
functions of the final neutrino energy $E_\nu^\prime$, for the initial
neutrino energy $E_{\nu}=3T$ (red dotted lines in
Fig. \ref{fig:neutrino-rates-comparison}).

As anticipated, Sky3s gives different scattering rate and average
scattering angle from Sky3 at the same $E_{\nu}^{\prime}$. For most
values of $E_{\nu}^{\prime}$, Sky3s gives smaller neutrino scattering
rates and average scattering angles (larger average cosine of the
scattering angle) than Sky3. These can affect simulations of
proto-neutron star evolution and supernova explosion. However, to that
purpose, it would be necessary to compute tables of moments of the
angular distribution of scattering rates \cite{Duan2023} and include
those into the neutrino transport of the supernova simulation. This
goes far beyond the scope of this article and is left for future work.

\section{Conclusion} \label{sec:conclusion}
This work aims to determine the spin-dependent terms of the new Skyrme
functionals introduced in Ref. \cite{Duan2024} to obtain realistic
values for the Landau parameters $G_0$ and $G'_0$, and thus a
reasonable description of the spin-related properties of nuclear
matter.

Since it is not easy to get precise information on spin dependent
terms from experiment, especially for neutron-rich matter at high
density, we have fitted the values of the Landau parameters and the
effective-mass splittings in spin polarized matter for Sky3s and Sky4s
to microscopic BHF results. In this way, most of the parameters of the
spin-dependent parts of the two new interactions have been
adjusted. This is in the same spirit as what was done in
Ref. \cite{Duan2024} for the spin-independent terms by fitting BHF
results for the neutron-matter EoS and the effective masses.

However, while the spin-independent gradient terms were constrained in
Ref. \cite{Duan2024} from the fit to finite nuclei, we have no
information about the spin-dependent gradient terms and we chose to
set them equal to zero. These terms of the functional should be
refined in future studies.

Now, the response functions can be computed with the new Skyrme
functionals having realistic effective masses and Landau parameters.
Since the typical momentum transfers in neutrino scattering processes
in (proto-)neutron stars are not very high, this is sufficient to get
reasonable neutrino rates, while the still undetermined spin-dependent
gradient terms contribute only at higher momentum transfers.

Concerning uniform matter properties, the new functionals do of course
not provide more predictivity than the BHF calculations to which they
were fitted. But they are much easier to use, in particular, for the
computation of the EoS and of the composition of neutron-star matter
in $\beta$ equilibrium, for finite temperature calculations, response
functions, and neutrino rates. Furthermore, since they were also
fitted to finite nuclei, they should also be suitable for the
description of the neutron-star crust. Therefore, the two new
interactions Sky3s and Sky4s will be useful to continue towards the
objective of Ref. \cite{Duan2023} to compute neutrino rates for
astrophysical simulations. More generally, they can be used in all
kind of future studies in nuclear astrophysics where simple Skyrme
energy functionals are needed because other approaches are
computationally too costly.

It should be mentioned, however, that the old BHF results for the
Landau parameters of
Refs. \cite{Zuo2003PNM,Zuo2003SNM,Zuo2003communications} that we used
for our fits have some problems. For instance, the results for $G_0$
in symmetric nuclear matter shown in Refs. \cite{Zuo2003SNM} do not
agree with those of Ref. \cite{Zuo2003communications}.  And in
addition to the problem discussed after Eq.~\eqref{eq:G0-chi}, it
should be mentioned that using those values for $G_0$ in pure neutron
matter, the energy of fully spin polarized neutron matter is too high
compared to the results of \cite{Tews2020,Vidana2016} as seen in
Fig. \ref{fig:EoS-PNM-polarization-nopolarization}. Therefore, it
would be important that the question of spin Landau parameters be
revisited by some new microscopic calculations.
\begin{acknowledgments}
We thank M. Bigdeli, G. H. Bordbar, and I. Vida\~na for sending us
their unpublished results for the effective masses in spin-polarized
matter. Mingya Duan is grateful for the support of the China
Scholarship Council (CSC No. 202006660002).
\end{acknowledgments}


\bibliography{references}

\end{document}